\documentclass[useAMS,usenatbib]{mn2e}
\usepackage{times}
\usepackage{natbib}
\usepackage{graphicx}
\usepackage{amssymb}
\usepackage{epsfig}

\title[Number counts and clustering of bright DRGs in the UDS
EDR]{Number counts and clustering properties of bright Distant Red
  Galaxies in the UKIDSS Ultra Deep Survey Early Data Release}
\author[S. Foucaud et al.]{S.  Foucaud$^{1}$\thanks{E-mail:
    Sebastien.Foucaud@nottingham.ac.uk}, O.  Almaini$^{1}$, I.
  Smail$^{2}$, C.~J.  Conselice$^{1}$, K.~P.  Lane$^{1}$, A.~C.
  Edge$^{2}$, \newauthor C.  Simpson$^{3}$, J.~S.  Dunlop$^{4}$, R.~J.
  McLure$^4$, M. Cirasuolo$^4$, P. Hirst$^{5}$, M.~G.  Watson$^{6}$,
  \newauthor M.~J.  Page$^{7}$
  \\
  $^{1}$School of Physics and Astronomy, University of
  Nottingham, University Park, Nottingham NG7 2RD  \\
  $^{2}$Institute for Computational Cosmology, Department of
  Physics, Durham University, Durham DH1 3LE \\
  $^{3}$Astrophysics Research Institute, Liverpool John Moores University, Egerton Wharf, Birkenhead CH41 1LD \\
  $^{4}$SUPA\thanks{Scottish Universities Physics Alliance} Institute for Astronomy, University of Edinburgh, Royal Observatory, Edinburgh EH9 3HJ \\
  $^{5}$Joint Astronomy Centre, 660 N. A'ohoku Place, University Park,
  Hilo, Hawaii 96720, U.S.A.\\
  $^{6}$X-ray Astronomy Group, Department of Physics and Astronomy,
  University of Leicester, Leicester LE1 7RH \\
  $^{7}$Mullard Space Science Laboratory, University College London, Holmbury S. Mary, Dorking, Surrey RH5 6NT \\
}

\begin{document}

\def\comseb#1{{\bf #1}}

\date{Accepted ... Received ...; in original form ...}

\pagerange{\pageref{firstpage}--\pageref{lastpage}} \pubyear{2006}

\maketitle

\label{firstpage}

\begin{abstract}
  { We describe the number counts and spatial distribution of 239
    Distant Red Galaxies (DRGs), selected from the Early Data Release
    of the UKIDSS Ultra Deep Survey. The DRGs are identified by their
    very red infrared colours with $(J-K)_{AB}\!>\!1.3$, selected over
    0.62 deg$^2$ to a $90\%$ completeness limit of $K_{AB}\!\simeq\!
    20.7$.  This is the first time a large sample of bright DRGs has
    been studied within a contiguous area, and we provide the first
    measurements of their number counts and clustering.  The
    population shows strong angular clustering, intermediate between
    those of $K$-selected field galaxies and optical/infrared-selected
    Extremely Red Galaxies.  Adopting the redshift distributions
    determined from other recent studies, we infer a high correlation
    length of $r_0\!\sim\!11\,h^{-1}$Mpc. Such strong clustering could
    imply that our galaxies are hosted by very massive dark matter
    halos, consistent with the progenitors of present-day $L\gtrsim
    L_*$ elliptical galaxies}.
\end{abstract}


\section{Introduction}
\label{sec:intro}

A new near-infrared selection technique has been developed in recent
years to sample galaxies in the high-redshift Universe.  By relying on
purely near-infrared colours, this potentially avoids many biases
which are inherent in optical techniques, particularly for detected
dusty and/or evolved galaxies.  \cite{2003ApJ...587L..79F} argue that
the simple $(J-K)_{AB}\!>\!1.3$ colour selection criteria produces a
sample that is mainly populated by galaxies at $z\!>\!2$, at least at
faint $K$-band magnitudes ($K_{AB}\!\gtrsim\!21$). These are the
so-called Distant Red Galaxies (hereafter DRGs).  In the Faint
Infrared Extragalactic Survey (FIRES), \cite{2003ApJ...587L..79F}
selected 14 candidate galaxies at $z\!>\!2$ to a depth of
$K_{AB}\!<\!24.4$, of which 6 were spectroscopically confirmed
\citep{2003ApJ...587L..83V}.  \cite{2005ApJ...624L..81L} found that
approximately $70\%$ of DRGs are dusty star forming galaxies and the
remaining $30\%$ are passively evolved galaxies.

Work by \cite{2003ApJ...599..847R} suggest that DRGs may be a
significant constituent of the $z\!\sim\!2-3$ universe in terms of
stellar mass.  \cite{2004ApJ...616...40F} demonstrated that the
average rest-frame optical colours of DRGs fall within the range
covered by normal galaxies locally, unlike the Lyman-break galaxies
(LBGs -- \citeauthor{1996AJ....112..352S}
\citeyear{1996AJ....112..352S}) which are typically much bluer.
Larger samples of DRGs covering a wide range in stellar mass are now
required to fully understand the importance of this population. In
particular, studies conducted so far have (by necessity) concentrated
on DRGs selected over relatively small areas, and very little is known
about the bright end of this population.

In terms of stellar mass, metalicity and star formation rate
\cite{2005ApJ...633..748R} find strong similarities between
optically-selected and near-infrared selected galaxy samples.
Clustering offers an alternative way to to study these populations. At
large scales, the galaxy distribution is dominated by dark mater halo
clustering, which is a strong function of halo mass.  Several studies
have measured strong clustering strength for high redshift galaxies
selected in the near-infrared ($r_0\!=\!10-15h^{-1}$Mpc)
\citep{2003ApJ...588...50D,DRG_MUSIC_astroph}, comparable to the most
luminous galaxies in the local universe.

In this paper we present a study of the first large sample of DRGs
selected at bright infrared magnitudes ($K_{AB}\!<\!21$) in a
contiguous area.  We analyse the number counts and clustering and draw
conclusions about their likely origin.  Throughout this paper, we
assume $\Omega_m\!=\!0.3$, $\Omega_{\Lambda}\!=\!0.7$ and
$h\!=\!H_0/70$~km~s$^{-1}$~Mpc$^{-1}$.

\section[]{UKIDSS UDS Early Data Release}
\label{sec:uds}

\subsection{Survey and Early Data Release}
\label{subsec:uds-edr}

The UKIRT Infrared Deep Sky Survey (UKIDSS --
\citeauthor{UKIDSS_astroph}
\citeyear{UKIDSS_astroph})\footnote{\texttt http://www.ukidss.org} is
began observations in Spring 2005, using the Wide-Field Camera (WFCAM
-- Casali et al., in prep.) at the 3.8-m United-Kingdom InfraRed
Telescope (UKIRT).  Comprising 5 sub-surveys, it will take 7 years to
complete and will cover a range of areas and depths.  The deepest of
these 5 sub-surveys, the Ultra-Deep Survey (UDS) aims to cover 0.8
deg$^2$ to a depth of $K_{AB}\!=\!25.0$, $H_{AB}\!=\!25.4$,
$J_{AB}\!=\!26.0$ ($5\sigma$, point-source).  It is centred on the
Subaru/XMM-Newton Deep Survey field (SXDS - \citeauthor{sxds}
\citeyear{sxds}) at $02^h18^m00^s$, $-05^{\circ}00'00''$ (J2000).

Since February 10 2006, the UKIDSS Early Data Release (EDR) has been
available to the ESO community\footnote{\texttt
  http://surveys.roe.ac.uk/wsa}. A full description of this data
release is given in \cite{UKIDSS-EDR_astroph}.

\subsection{Image stacking and mosaics}
\label{subsec:uds-mos}

The stacking of the UDS EDR data was performed by our team using a
slightly different recipe to the standard UKIDSS pipeline. Given the
(relatively) small size of the field, it is possible to create a full
mosaic before extracting catalogues, rather than merge catalogues
extracted from individual chips. Given the various jitter and offset
sequences, this helps to optimise depth in overlap regions and to
produce a more homogeneous final image.

Each observation block consists of a 4-pt mosaic to tile the $0.8$
deg$^2$ field, producing 16 images each of $\simeq\!15$ minutes
exposure (see \citeauthor{UKIDSS-EDR_astroph}
\citeyear{UKIDSS-EDR_astroph}). Individual reduced frames for each
observation block were extracted from the UKIDSS pipeline as the
starting point for our final mosaiced stack (including astrometric and
photometric solutions).  We used the variance map produced by the
pipeline to weight each frame before stacking and rescaled the pixel
flux for each individual image using the pipeline zero-points.  The
stacking was carried out in a two-step process by using the
\texttt{SWarp} software, an image resampling tool \citep{swarp}.  The
final mosaiced images in the $J$- and $K$-band have the same pixel
scale of $0.1342''$, with identical field centres and image scales to
simplify catalogue extraction.  The resulting images were visually
inspected, and bad regions were masked out (areas around saturated
stars, cosmetic problem areas, and low signal-to-noise borders).
After masking, the usable area of this frame with uniform coverage is
$0.62$ deg$^2$.  We note that this is smaller than the expected $0.8$
deg$^2$ mosaic since the UDS field centre was moved by $\sim\!8$
arcmin shortly after observations began (to allow the use of brighter
guide stars).

The image seeing measured from the mosaiced images is $0.69''$ FWHM
in $K$ and $0.80''$ in $J$.  The RMS accuracy of our astrometry is
$\simeq\!0.05''$ (i.e. $<\!1$ pixel) and for our photometry
the RMS accuracy is $\lesssim\!2\%$ in both $J$- and $K$- bands
\citep{UKIDSS-EDR_astroph}. From direct measurements of noise in a
$2''$ aperture on the image we estimate $5\sigma$ limiting magnitudes
of $K_{AB}\!=\!22.5$ and $J_{AB}\!=\!22.5$.

\subsection{Catalogue extraction}
\label{subsec:uds-cats}

We found that the standard UKIDSS source detection software did not produce
optimal catalogues for the UDS. We therefore produced a much improved
catalogue for the EDR by using the \texttt{SExtractor} software
\citep{sextractor}. The $K$-band image was used as the source detection image,
since this is measurably deeper for most galaxy colours. All $K_{AB}$
magnitudes quoted below are total magnitudes extracted using the
\texttt{SExtractor} parameter \texttt{MAG\_AUTO}, while all colour measurements
are obtained from fixed $2''$ aperture magnitudes.

To optimise our catalogue extraction we performed a series of
simulations to fine tune the \texttt{SExtractor} parameters.
Artificial point-like sources were added to the real $K$-band image
using the observed PSF with FWHM $=\!0.69''$ (rejecting regions
containing bright sources), and distributed with magnitudes in the
range $14\!<\!K_{AB}\!<\!24$.  From the resulting new image a catalogue
was extracted using \texttt{SExtractor} and compared with the list of
artificial source positions. This process was repeated 1000 times, and
the resulting statistics allow us to estimate the catalogue
completeness and the evolution of photometric errors.  Using these
simulations, \texttt{SExtractor} detection parameters were tuned to
maximise completeness at the noise-determined $5\sigma$ depth of
$K_{AB}\!=\!22.5$, while simultaneously minimising the number of spurious
sources.  While formally optimised for point-like sources, we note
that these were close to optimal when we generated artificial sources
using a substantially more extended PSF (FWHM $\!=\!1.2''$).

Using these parameters, we extracted $78709$ sources over $0.62$deg$^2$
  from the image, of which $34098$ were determined to be
  unsaturated, unmasked and from  regions of uniform coverage
  to  $K_{AB}\!<\!22.5$. These form the basis of the analysis outlined
  below.
Assuming the background noise is symmetric about zero, we can estimate the
spurious fraction by extracting sources from the inverted image and comparing
with the number of sources extracted from the normal image. At our
magnitude limit of $K_{AB}\!=\!22.5$ the fraction of spurious
detections is found to be less than $1\%$, while the completeness level is
above $70 \%$ (for point sources).

\section[]{Selection and number counts}
\label{sec:drg}

\subsection{Selection of DRGs}
\label{subsec:drg-sel}

\begin{figure}
\begin{center}
  \resizebox{\hsize}{!}{\includegraphics{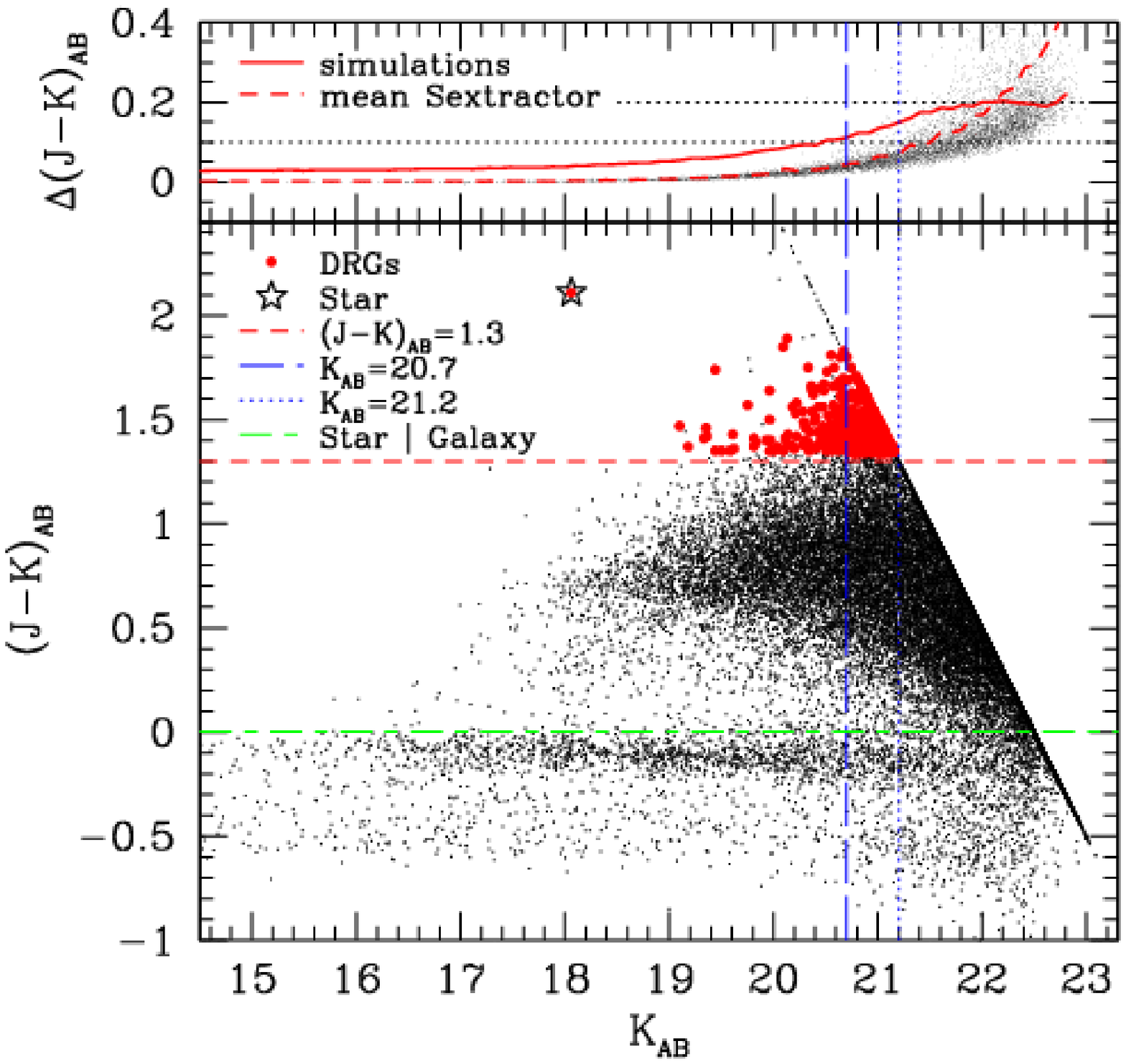}}
  \caption{{\it Lower panel:} $(J-K)_{AB}$ against $K_{AB}$ for
   sources in the UDS EDR field. Small points
    represent the full population of K-band selected objects, while
    larger points are those selected with $(J-K)_{AB}\!>\!1.3$ (and
    visually confirmed).  The  colour criteria, the magnitude
    limit at $K_{AB}\!=\!21.2$ and the magnitude completeness at
    $K_{AB}\!=\!20.7$ are highlighted, as is the crude boundary between
    galaxies and the stellar locus at $(J-K)_{AB}\!=\!0$.  {\it Upper
      panel:} Errors as derived by \texttt{SExtractor} on the
    $(J-K)_{AB}$ colours (for display purposes only $1/5$th of the
    points are shown). Mean values are also displayed, as are errors
    derived from simulations.  }
  \label{fig:sel}
\end{center}
\end{figure}

From the catalogue described above we selected objects using the
$(J-K)_{AB}\!>\!1.3$ criteria. A visual inspection of each source was then
required to remove spurious detections, which at these extreme colours
was found to be a relatively large fraction ($\sim\!20\%$). The majority
are caused by diffraction spikes and cross-talk images
\citep{UKIDSS-EDR_astroph} and are easy to identify and reject. This
leaves 369 DRGs at $K_{AB}\!<\!21.2$, which represents the largest sample
selected over a contiguous area. The surface density derived
  is $n\!=\!0.163\pm0.009$ arcmin$^{-2}$.  Figure~\ref{fig:sel} shows
the $(J-K)$ colour of these galaxies versus $K$-band magnitude.  The
object shown by a star was classified as a point-like source in our
global catalogue, and is confirmed to be a star after visual
inspection.

\subsection{Photometric errors and contamination of DRG sample}
\label{subsec:drg-cont}

Since our sample is based on $(J-K)$ colour selection it is vital to
carefully consider the effects of photometric errors. Since most
galaxies show substantially bluer colours (Figure~\ref{fig:sel}) we
can expect the number density of DRGs to be artificially boosted at
fainter magnitudes, as errors push objects above the $(J-K)_{AB}\!>\!1.3$
selection boundary.  As a lower limit to this contamination we could
use the photometric errors derived from \texttt{SExtractor}, and these
are shown as a function of magnitude in Figure~\ref{fig:sel}. Our
experience suggests that analytically-determined errors from
\texttt{SExtractor} are likely to be underestimates, so we use the
mean photometric errors obtained from the simulations described in
section~\ref{subsec:uds-cats}.

We used our simulated errors to estimate the contamination by
randomising the real galaxy catalogue using Monte-Carlo simulations.
For each object in our full catalogue, we allow the $(J-K)$ colour to
vary assuming a Gaussian distribution with a standard deviation
corresponding to the chosen photometric error. We then re-select our
catalogue using the $(J-K)_{AB}\!>\!1.3$ criteria, and repeat this process
1000 times. This should provide an approximate upper limit on
contamination, since we are randomising the observed galaxy catalogue
(which already suffers from the effects of photometric errors).

Defining the contamination fraction from the number of objects
scattered into our selection boundary minus those which are scattered
out, our simulated source errors yield contamination fractions of
$(46.0\pm3.8)\%$ at the limiting magnitude of $K_{AB}\!<\!21.2$, falling
to $(16.8\pm3.6)\%$ at $K_{AB}\!<\!20.7$ (the estimated completeness
limit; see Section~\ref{subsec:drg-counts}).  We note that the typical
error on the colour is $\Delta(J-K)_{AB}\!\sim\!0.1$ at $K_{AB}\!=\!20.7$.  As
shown in \cite{AEGIS_DRG_astroph}, a slight change of the $(J-K)$
colour selection does not have a major affect on the redshift
distribution.

Based on these values we conclude that our number counts and
clustering measurements are reasonably robust at $K_{AB}\!<\!20.7$, but
will become increasingly unreliable at fainter magnitudes. We will
therefore adopt a limit of $K_{AB}\!=\!20.7$ for further study, which
produces a sample of $239$ DRGs.

\subsection{Number counts of DRGs}
\label{subsec:drg-counts}

\begin{figure}
\begin{center}
  \resizebox{\hsize}{!}{\includegraphics{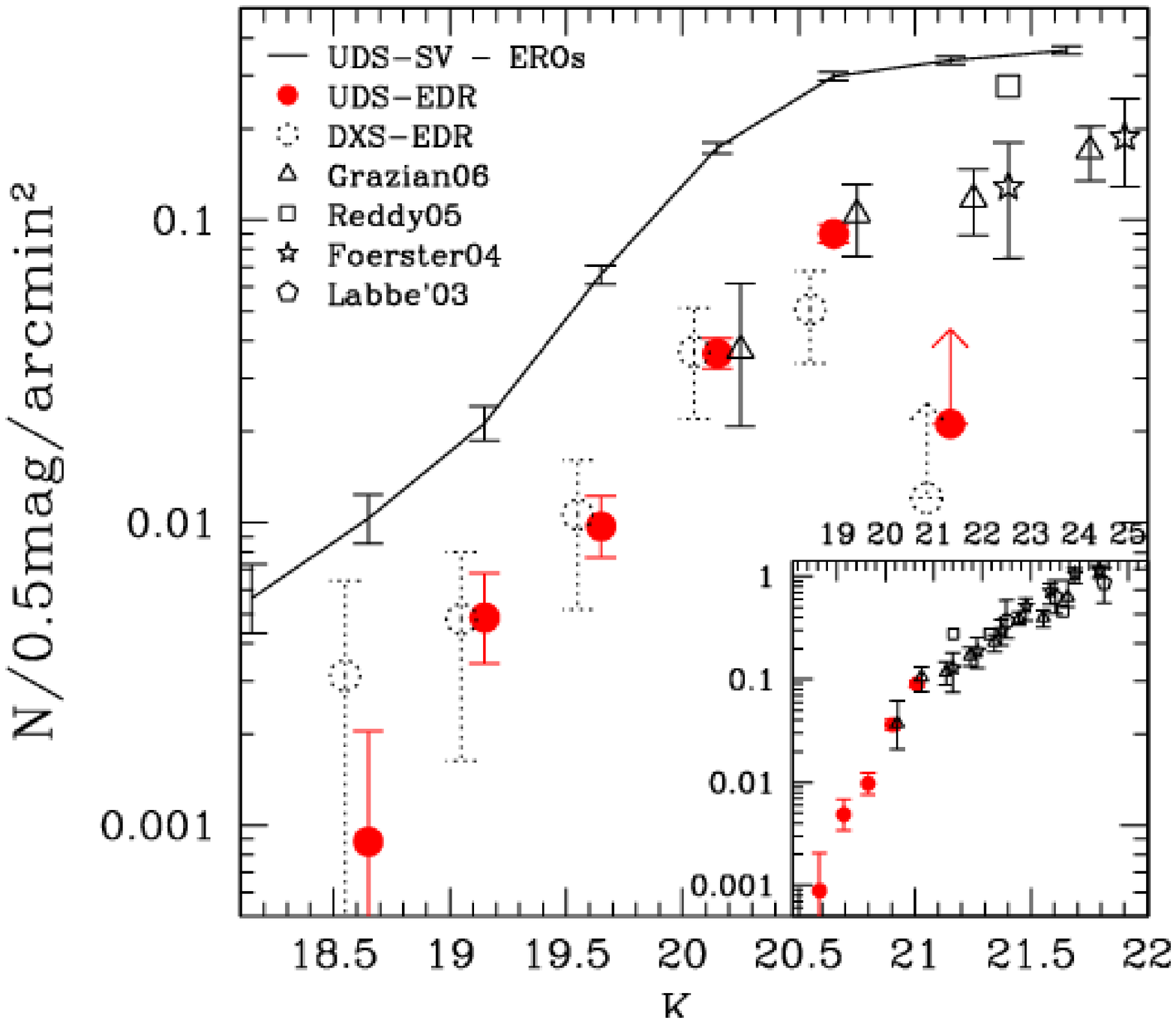}}
  \caption{$K$-band differential number counts for our sample of DRGs.
    The errorbars plotted are computed from poissonian small number
    statistics \citep{1986ApJ...303..336G}.  For comparison, we have
    overplotted the number counts for DRGs derived from the DXS EDR
    survey, with errors representing the field-to-field variance (see
    section~\ref{subsec:drg-dxs} -- these points are slightly shifted
    for display purposes), and from the literature from fainter
    samples
    \citep{2003AJ....125.1107L,2004ApJ...616...40F,2005ApJ...633..748R,DRG_MUSIC_astroph}.
    EROs number counts from the UDS/DXS SV sample are shown as well,
    for comparison. The plot inset shows number counts over a larger
    magnitude range.}
  \label{fig:counts}
\end{center}
\end{figure}

Figure~\ref{fig:counts} shows the $K$-band differential number counts
of our sample of DRGs.  The number counts indicate that our sample is
complete up to approximately $K_{AB}\!\simeq\!20.7$, after which the
counts are clearly dropping. This defines our estimated completeness
limit.  Our simulations suggest that the contamination due to
photometric errors will be $\sim\!16\%$ at $K_{AB}\!<\!20.7$. We
conclude the dominant source of error in our number counts will
be Poisson counting errors (plotted) and cosmic variance (discussed in
section~\ref{subsec:drg-dxs}).

At bright magnitudes (e.g.  $K_{AB}\!<\!20$) our UKIDSS data are entirely
unique, and no studies exist in the literature for comparison.  At
fainter magnitudes, our counts are in very good agreement with the DRG
counts from the \cite{DRG_MUSIC_astroph} sample. They are also
consistent with the number counts from the AEGIS survey (Foucaud et
al. in prep.; see also \citeauthor{AEGIS_DRG_astroph}
\citeyear{AEGIS_DRG_astroph}).  Combining literature data with the
present work we can examine the global shape of the DRG number counts
over a very wide dynamic range ($18.5\!<\!K_{AB}\!<\!25.0$).  This strongly
suggests a break feature in the slope at $K_{AB}\!\sim\!20.5$ which is an
effect already seen in the global K-band number counts (e.g.
\citeauthor{GCW} \citeyear{GCW}).

We note that the projected density of DRGs is
approximately 10 times lower than EROs, and approximately 100 times
lower than the global galaxy counts at a given magnitude.

\subsection{Cosmic variance}
\label{subsec:drg-dxs}

As a simple test of cosmic variance, and to investigate whether the
UDS is an unusual field, we used the data available from the UKIDSS
Deep Extragalactic Survey (DXS) to perform a comparison study.  The
DXS is the other deep extragalactic component of UKIDSS
\citep{UKIDSS_astroph}, consisting of 4 fields with a 7-year goal of
observing 35 deg$^2$ to depths of $K_{AB}\!=\!22.7$ and
$J_{AB}\!=\!23.2$. We used data from 3 fields observed in both $J$-
and $K$-bands in the UKIDSS EDR, covering $\sim\!2900$ arcmin$^2$,
$\sim\!4500$ arcmin$^2$ and $\sim\!2900$ arcmin$^2$ respectively.
While exposure times are similar to the UDS EDR, the observing
conditions are generally poorer. Direct comparison is also complicated
by the different source extraction methods used by the DXS.

We applied the same selection method described in
section~\ref{subsec:drg-sel}, except that we did not visually inspect
the samples. This selects $1523$ objects in total, of which we
estimate approximately $20\%$ are likely to be spurious (see
section~\ref{subsec:drg-sel}), with a similar fraction likely to be
artificially boosted due to photometric errors at faint magnitudes
(section~\ref{subsec:drg-cont}). Since these errors are smaller than
the errors in the DXS counts, for simplicity we opt not to make these
corrections in our comparison with the UDS. We derived a median
surface density of $n\!=\!0.176\pm0.075$ arcmin$^{-2}$ at
$K_{AB}\!=\!21.2$, in very good agreement with the UDS value.

The resulting median counts from the 3 DXS samples are overplotted in
figure~\ref{fig:counts}, with errors representing the field-to-field
RMS variance. The agreement with UDS is very good.
Although no corrections were applied, this crude comparison suggests
that the density of DRGs is stable and broadly consistent between
fields.

\section{The clustering of DRGs}

\subsection{Angular clustering}
\label{subsec:drg-cf}

\begin{figure}
\begin{center}
  \resizebox{\hsize}{!}{\includegraphics{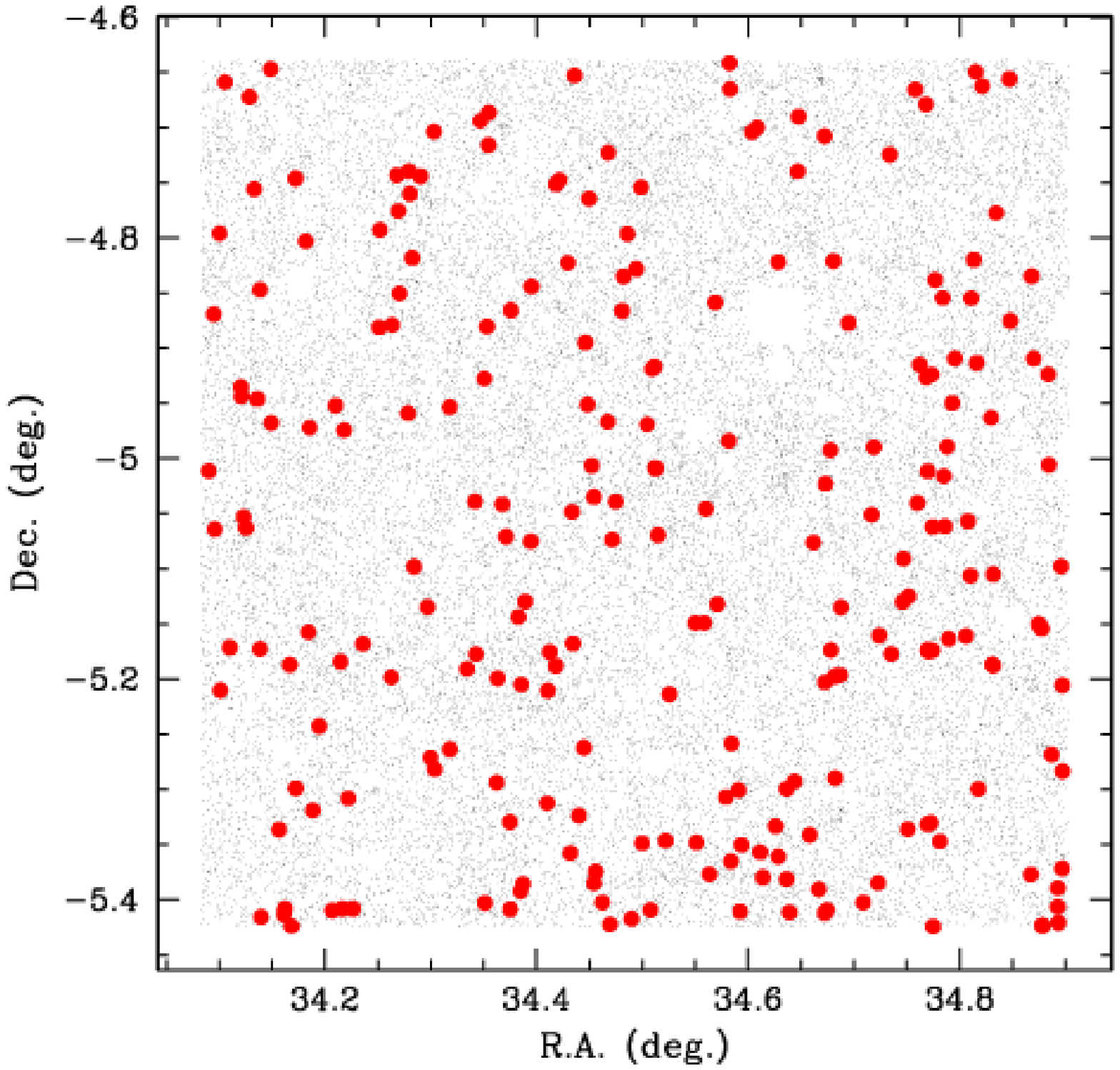}}
  \caption{Distribution of our sample of DRGs on the sky, providing a visual impression of the large scale structure.  Small dots represent the full K-band galaxy sample, while larger symbols are the DRGs with
    $K_{AB}\!<\!20.7$.}
  \label{fig:xy}
\end{center}
\end{figure}

In Figure~\ref{fig:xy}, we display the distribution of our sample of
DRGs on the sky. Visually the DRGs appear strongly clustered.  As a
more quantitative measure, we evaluate the 2-point angular correlation
function, cutting our sample at $K_{AB}\!=\!20.7$ as before.

\begin{figure} 
\begin{center}
  \resizebox{\hsize}{!}{\includegraphics{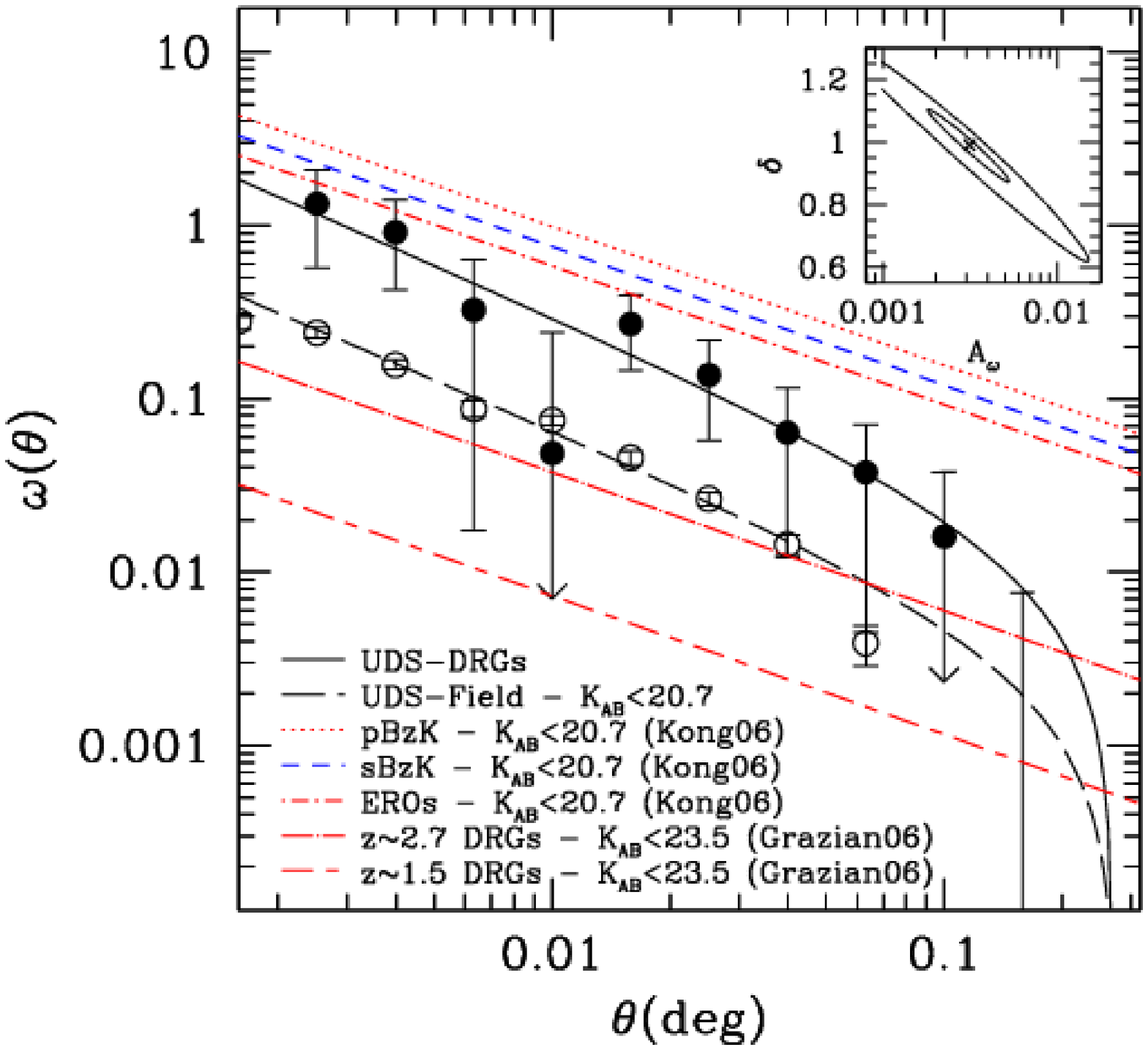}}
  \caption{2-point angular correlation function determined for our sample of
    DRGs (full circle) and for the field galaxies (open circles) with
    $K_{AB}\!<\!20.7$ and their fitted power-laws . The plot inset
    presents the $\chi^2$ minimisation to estimate the best-fitted
    values of $A_{\omega}$ and $\delta$ for our sample of DRGs, with
    contours showing $1\sigma$ and $3\sigma$ confidence levels. For
    comparison, measurements from the literature for different samples
    are overplotted, assuming a slope of $\delta\!=\!0.8$
    \citep{2006ApJ...638...72K,DRG_MUSIC_astroph}.}
  \label{fig:cf}
\end{center}
\end{figure}

To measure the angular correlation function $\omega(\theta)$ and
estimate the related poissonian errors we used the \cite{LS}
estimators.  Figure~\ref{fig:cf} shows the correlation function
derived from our DRG sample.  The best fit for the angular correlation
is assumed to be a power-law as:
$$\omega(\theta)\!=\!A_{\omega}(\theta^{-\delta}-C_{\delta})$$
with
the amplitude at 1 degree
$A_{\omega}\!=\!3.1^{+2.1}_{-1.3}\times10^{-3}$, the slope
$\delta\!=\!1.00\pm0.10$, and the integral constraint due to the
limited area of the survey $C_{\delta}\!=\!3.71$ (determined over the
unmasked area).

For comparison we also determine the angular correlation function for
the full K-selected sample to a limiting magnitude of
$K_{AB}\!<\!20.7$, as shown in figure~\ref{fig:cf}. The derived
amplitude at 1 degree is
$A_{\omega}\!=\!8.1^{+2.3}_{-1.8}\times10^{-4}$ for a slope of
$\delta\!=\!0.96\pm0.05$. Fixing the slope to the standard literature
value ($\delta\!=\!0.8$) we derived an amplitude of
$A_{\omega}\!=\!1.8^{+0.1}_{-0.1}\times10^{-3}$ which is in good
agreement with the work of \cite{2006ApJ...638...72K} to the same
magnitude limit.

As shown in figure~\ref{fig:cf}, by comparing with measurements from
\cite{2006ApJ...638...72K} at the same magnitude limit of
$K_{AB}\!<\!20.7$, the amplitude of the DRG sample is $\sim\!5$ times
higher than that of field galaxies but a factor of $\sim\!2$ lower
than EROs and BzK-selected galaxies. Interestingly, our derived
clustering amplitude is notably higher than measurements for DRGs
obtained by \cite{DRG_MUSIC_astroph}.  This issue is explored further
below.


\subsection{Spatial correlation lengths and biasing}
\label{sec:concl}

In order to compare the clustering of galaxy populations at different
redshifts we must derive spatial correlation measurements.  Assuming a
redshift distribution for our sample we can derive the correlation
length and a linear bias estimation using the relativistic Limber
equation \citep{2000MNRAS.314..546M}.  The difficulty here is to have
a realistic estimation of the redshift distribution of our sample.
While \cite{2003ApJ...587L..79F} have designed the
$(J-K)_{AB}\!>\!1.3$ criteria to select $z\!>\!2$ galaxies,
\cite{DRG_MUSIC_astroph} and \cite{AEGIS_DRG_astroph} have shown that
the redshift distribution is broad, and the fraction of $z\!<\!2$
galaxies increases at bright magnitudes.

As a preliminary investigation we assume a Gaussian form for the
redshift distribution, fixing the mean and standard deviation to match
recent studies of fainter DRGs.  Analytical and fitted redshift
distributions make no significant difference to our derived
correlation lengths.  The correlation length can then be evaluated if
we fix the slope of the correlation function to
$\gamma\!=\!1+\delta\!=\!2.0$.  In the DRG sample of
\cite{DRG_MUSIC_astroph}, a mean redshift of $\bar{z}\!=\!1.5$ was
found at a limit of $K_{AB}\!<\!22.0$. Using this mean redshift with
$\sigma\!=\!0.5$ (broadly matching their distribution) we determine
$r_0\!=\!14.1^{+4.8}_{-2.9}\,h^{-1}$Mpc and $b\!=\!4.8^{+1.7}_{-1.1}$
for our DRGs. \cite{AEGIS_DRG_astroph} suggest that very bright DRGs
lie at even lower redshifts. If we adopt the observed redshift
distribution of their spectroscopic sample we derive
$r_0\!=\!11.0^{+3.7}_{-2.3}\,h^{-1}$Mpc and $b\!=\!4.0^{+1.4}_{-0.8}$,
when the approximated gaussian form gives (with $\bar{z}\!=\!1.0$ and
$\sigma\!=\!0.25$) $r_0\!=\!11.1^{+3.8}_{-2.3}\,h^{-1}$Mpc and
$b\!=\!4.0^{+1.4}_{-0.8}$.

\cite{DRG_MUSIC_astroph} also split their sample according to
redshift, and found $r_0\!=\!7.4^{+3.5}_{-4.9}\,h^{-1}$Mpc for DRGs at
$1\!<\!z\!<\!2$ with $K_{AB}\!<\!22$. Our sample is substantially
brighter, so, even if the two samples are probing slightly different
redshifts, our larger correlation length could be interpreted as
evidence for luminosity segregation, e.g. we may be sampling more
luminous galaxies which are more strongly biased
\citep{2006A&A...451..409P}.  We note that \cite{Quadri_astroph} find
no relationship between K-magnitude and correlation length for
K-selected galaxies, but instead find a strong relationship between
colour and correlation length (with redder galaxies being more
strongly clustered). This is not inconsistent with our findings, since
we are studying segregation {\em within} a colour-selected sample.

As shown in figure~\ref{fig:cf}, our measurement is comparable to the
clustering of EROs or $BzK$ selected galaxies at the same limiting
magnitude, which are likely to sample a similar redshift distribution
to our DRGs, and hence show similar correlation lengths.
\cite{2003A&A...405...53O} compared the clustering of EROs with
predictions of Dark Matter halo models. Adopting the same technique,
we can infer that bright DRGs are likely to be hosted by very massive
dark matter halos (possibly $M\!>\!10^{14}M_{\odot}$), consistent with
the potential progenitors of present-day $L\!\gtrsim\!L_*$ elliptical
galaxies.

Finally we note that while the slope of our angular correlation
function is steep, the fiducial value of $\delta\!=\!0.8$ is only
excluded by our sample at a $1\sigma$-level. Further data will be
required to investigate whether this may be caused by a possible
transition between inner and outer halo occupation numbers (e.g.
\citeauthor{2005ApJ...635L.117O} \citeyear{2005ApJ...635L.117O}).

\section{Summary}
\label{sec:summ}

We have extracted a large sample of bright Distant Red Galaxies from
the UKIDSS UDS EDR. Our catalogue contains $369$ DRGs to a limiting
magnitude of $K_{AB}\!=\!21.2$, extracted over an area of $0.62$
deg$^2$.  The fainter $K_{AB}\!>\!20.0$ number counts are in good
agreement with previous estimates, while at brighter magnitudes the
sample is unique.  Using simulations we determined that contamination
due to photometric errors is below $\sim\!16\%$ at an approximate
completeness limit of $K_{AB}\!<\!20.7$.

From this sample we extracted a sub-sample of 239 bright DRGs to a
limit of $K_{AB}\!=\!20.7$. These bright DRGs appear highly clustered,
and we determine a correlation length of $r_0\!\simeq\!11 \,h^{-1}$Mpc
and a bias measurement $b\!\simeq\!4.5$, assuming the sample lies at a
mean redshift of $\bar{z}\!=\!1.0$ with a standard deviation of
$\sigma\!=\!0.25$ (consistent with studies at similar depths --
\citeauthor{AEGIS_DRG_astroph} \citeyear{AEGIS_DRG_astroph}). They
appear more clustered than fainter samples of DRGs derived at these
redshifts, which may be evidence for luminosity segregation, in
agreement with biased galaxy formation scenarios.

\section*{Acknowledgements}

The authors would like to thank the referee for a detailed and
thorough report which improved the paper, Stephen Warren and Stephen
Serjeant for their useful comments, and Nigel Hambly and Nick Cross
for their help with data reduction. We are also grateful to the staff
at UKIRT for making these observations possible, and acknowledge the
Cambridge Astronomical Survey Unit and the Wide Field Astronomy Unit
in Edinburgh for processing the UKIDSS data.  SF, CS and MC
acknowledge funding from PPARC. OA, IS and RJM acknowledge the support
of the Royal Society.


\label{lastpage}

\end{document}